# Mars excitement in Australian newspapers, 1877–1899: Humour and the public negotiation of astronomical knowledge

**Richard de Grijs**
School of Mathematical and Physical Sciences, Macquarie University, Sydney, Australia.

**Abstract:** *Speculation about Martian canals was a recurring feature of late nineteenth-century popular astronomy. This paper examines how colonial newspapers used humour to negotiate the epistemic uncertainty and interpretive excess associated with canal theory. Drawing on over one thousand metropolitan and regional Australian newspapers published between 1877 and 1899, we identify five overlapping modes of humour: imported metropolitan wit; satire of modern engineering culture; humour grounded in observational uncertainty; scale-based exaggeration and colonial self-comparison; and overt sceptical parody. These modes tracked shifting relationships between observation, interpretation and authority, allowing newspapers to entertain speculative ideas while marking the limits of scientific credibility. At the same time, humorous treatments positioned Australian readers within a global culture of science and modernity. Comparisons with projects such as the Suez and Panama Canals, and with European and American astronomers, aligned colonial audiences with metropolitan discourse, even as local experience with land, water and scale shaped the tone of satire. We demonstrate that Australian newspapers did not passively transmit overseas ideas but actively reworked them through humour, balancing fascination with restraint. More generally, this case suggests that humour could function as a cultural strategy through which historical newspaper audiences engaged with speculative scientific claims without abandoning trust in scientific authority.*

## Introduction

During the final decades of the nineteenth century, Mars became one of the most culturally charged objects in popular astronomy. Periodic oppositions,[1] improvements in telescopic resolving power and the controversial interpretation of newly observed surface markings as 'canals' combined to elevate the planet from a routine observation target to a site of intense public interest. Newspapers played a central mediating role, translating technical observations into accessible narratives and embedding them within wider cultural frameworks that included engineering ambition, the planet's evolutionary decline and the possibility of intelligent alien life. Whereas the scientific context of this episode has been examined extensively (e.g., Crossley, 2011; Crowe, 1986; Lane, 2011; Sheehan, 1996; see also de Grijs, 2026a), the cultural modes through which Mars was rendered meaningful to readers—especially tone, irony and humour—have received comparatively little sustained analytical attention.

In this paper I focus on humour as an important mediating device in newspaper representations of "Mars excitement". Rather than treating humorous passages as marginal or incidental, I argue that humour functioned as a culturally productive strategy through which newspapers managed uncertainty, negotiated authority and regulated the boundaries between science and speculation or fantasy. Jokes, irony, parody and exaggeration were not simply ornamental; they shaped how readers were invited to understand what was known, what was contested and what might be safely doubted, or at least questioned.

The analysis draws on a systematically compiled collection of Australian newspaper articles published between the 1870s and 1899. They were assembled through targeted keyword searches of digitised colonial newspapers (including "Mars", "canals", "planet", "inhabited", "astronomers" and related terms) in the National Library of Australia's *Trove* database (https://trove.nla.gov.au/). The resulting dataset comprises more than one thousand items drawn from metropolitan[2] and regional newspapers across multiple Australian colonies (de Grijs, 2026a,b).[3] It includes original reporting, editorials, reprinted overseas articles and syndicated commentary; several hundred of which contain explicit humorous, ironic or satirical elements. Items were manually screened to identify instances where humour operated rhetorically rather than incidentally. Particular attention was paid to tone, metaphor, analogy and narrative framing, as well as to patterns of repetition and reprinting across titles.

---

[1] Oppositions are configurations in which Mars and the Sun are located on opposite sides of the Earth, bringing Mars to its closest approach and making it appear brighter and larger than usual in the night sky (see de Grijs, 2026a).

The period of interest is bracketed at early times by the discovery of the Martian satellites, Phobos and Deimos, in 1877; the database coverage ends in 1899, corresponding to the effective conclusion of the pre-fictional period in popular Mars reporting. As demonstrated in de Grijs (2026b), the article database is characterised by extensive reprinting of overseas material—particularly from British and American sources—combined with locally authored commentary, editorial framing and, occasionally, original observation. Australian newspapers were fully embedded within global telegraphic and syndication networks (e.g., Moyal, 1984; Potter, 2012), yet they exercised considerable discretion in how imported material was contextualised, presented, integrated or reinterpreted.

While the article collection necessarily reflects the biases of digitisation, optical-character-recognition (OCR) survival and editorial selection, its breadth across metropolitan and regional newspapers allows for comparative analysis of how humour functioned within different journalistic contexts. The five categories of humour identified below emerged through iterative reading and are not mutually exclusive; rather, they represent recurrent rhetorical modes through which Australian newspapers negotiated the epistemic and cultural implications of Martian canal speculation.

This makes the Australian press an especially revealing site for examining humour in popular astronomical discourse. Operating at a significant geographical and epistemic distance from the dominant metropolitan centres of scientific authority, Australian newspapers were dependent on overseas expertise yet rhetorically free to reinterpret, reframe and occasionally destabilise it. Humour offered a flexible means of participation: it allowed colonial readers to engage with international debates over Martian canals, habitability and planetary engineering while maintaining a degree of interpretive autonomy. Through wit and irony, newspapers could endorse or question scientific claims without directly challenging the legitimacy of the astronomical method.

Although this article focuses on Australian newspapers, the case also illuminates broader questions about how scientific uncertainty was negotiated in popular discussions of astronomy. Late nineteenth-century debates about Mars formed part of a wider international circulation of astronomical ideas, transmitted through telegraph networks and reprinted journalism. Examining how colonial newspapers incorporated these debates therefore offers insight into how speculative scientific claims were domesticated within local cultural contexts. The prominence of humour in Australian reporting suggests that comic framing could function as a cultural strategy for managing uncertainty about scientific claims without rejecting the authority of science itself.

Building on the structural analysis of circulation and attribution presented in de Grijs (2026b), this study shifts attention from *what* was reported to *how* it was reported. It identifies five recurring humorous modes that emerge sequentially across the period of interest: (1) inherited metropolitan humour; (2) explanatory wit; (3) speculative play; (4) exaggerated analogy; and (5) sceptical parody. These modes emerge sequentially but overlap in practice; they are not mutually exclusive, nor do they map neatly onto specific individuals or newspapers. Instead, they represent evolving cultural responses to the increasing interpretive instability of Mars as a scientific object. By emphasising humour as an analytical lens, this paper contributes to broader discussions in astronomy and culture as regards popular science, media mediation and the social management of uncertainty. I will demonstrate that humour was central to the late nineteenth-century cultural life of Mars. It enabled newspapers to sustain public interest while negotiating the limits of scientific credibility. In doing so, I position Australian journalism as an active participant in the cultural construction of planetary knowledge.

---

[2] Throughout this paper, the term *metropolitan* is used to denote the dominant centres of nineteenth-century scientific journalism and cultural authority—principally London, Paris and, increasingly, major American cities—from which astronomical news, commentary and humour circulated outwards to colonial newspapers.

[3] The full article collection can be accessed at https://astro-expat.info/Data/mars19c.html (for long-term preservation, see also
https://web.archive.org/web/20260112080309/https://astro-expat.info/Data/mars19c.html).

**Category 1: Humour as inherited metropolitan framing**

Before Australian newspapers developed distinctive humorous responses to Mars, they frequently reproduced humour already embedded in metropolitan reporting. Early jokes about Martian canals, irrigation and planetary engineering entered the Australian press mostly via reprinted British and American articles—particularly from sources such as the *Pall Mall Gazette*—and relied on international reference points including the Suez and Panama Canals. A strong example of this type of humour includes:

> The latest telescopic news from Mars is that the planet is one vast network of canals, a place, in fact (says the "Pall Mall Gazette") where all good Panama promoters might hope to go. (*Evening News*, 17 October 1898)

This line appeared across multiple Australian titles and is clearly identified as originating overseas. The humour depends on British readers' familiarity with Panama Canal scandals, not on any Australian reference point. Australian editors transmitted the joke unchanged, functioning as conduits rather than contributors. In this earliest category, humour was transmitted rather than generated: Australian newspapers functioned as distributors of metropolitan wit, embedding readers within a global culture of amused speculation without yet articulating a local voice. What was imported was not merely information but a tonal package—scientific authority laced with metropolitan irony.

A second illustrative example, taken from the German-language *Australische Zeitung* of 3 October 1882, explicitly reports that *Figaro* illustrations and commentary were being discussed in the *Revue*, including the claim—via the English astronomer Richard A. Proctor (1837–1888)—that Martian canals "… owe their origin to the industrious activity of the inhabitants of Mars", achievements that would rival even the construction of the Suez Canal. Here, humour and irony are already layered. This excerpt represents French visual satire, filtered through British interpretive commentary and then reproduced in an Australian newspaper. There is no attempt at localisation; Australia enters only as a receiver. This example shows that early humour was not just imported from British or American media but that it circulated through complex transnational media chains. This multi-stage circulation underscores that early humour was already internationalised well before reaching Australia; colonial newspapers did not inherit raw observation but a layered rhetorical artefact.

Several articles published in 1888 jokingly suggested that Martians had already duplicated the Suez Canal, completed a Panama Canal through their polar cap and, hence, outperformed Ferdinand Marie de Lesseps (1805–1894; developer of the Suez Canal) and European engineers. For example, in the *Argus* of 20 August 1888 we read,

> These hypothetical gentry have not only duplicated their Suez Canal before we have done anything more than merely talk about doing the same thing for ours, but have also succeeded in making their Panama Canal right through the polar ice cap, while the lottery loan that is going to put the finishing touch on the enterprise of M[onsieur]. de Lesseps has not yet been raised and now they are supposed to have carried out an irrigation work by the side of which the Mildura experiment and all the Chief Secretary's [of Victoria] pet schemes in addition look very small indeed.[4]

This humour relies primarily on contemporary European engineering failures and delays. The joke is meaningful, because readers recognise the Panama Canal project as a notorious international fiasco. The use of this type of humour supports the idea that early humour assumed metropolitan reference knowledge, was globally legible but not locally grounded and positioned Australian readers as observers of an international conversation.

Such material often carried an ironic tone, witty turns of phrase or playful speculation that originated outside Australia. When reproduced locally, this humour functioned less as editorial commentary than as part of the authoritative package of overseas science news. The presence of humour did not signal scepticism or distance; rather, it reflected the stylistic conventions of metropolitan scientific journalism. Nevertheless, in multiple cases, newspapers explicitly distanced themselves by attributing humorous excess to named overseas sources, e.g., "… says the *Pall Mall Gazette* …", "… according to the latest astronomical interviewer …" or "… has been confirmed … in England …" Australian editors thus allowed metropolitan irony to circulate without editorial endorsement or challenge. As such, humour functioned as a safe form of participation, allowing newspapers to entertain without assuming interpretive responsibility.

This category of humour is important because it establishes that humour entered Australian Mars reporting *with* the international discourse, not as a colonial distortion of it. Early Australian readers encountered Mars through a rhetoric already shaped by British and American media cultures, in which wit and scientific authority routinely coexisted within popular science journalism (e.g., Lightman 2007; Secord 2004). Only later did Australian newspapers begin to adapt humour to local contexts, concerns and sensibilities. Categories 2–5 therefore represent a gradual shift from inherited framing to locally inflected rhetorical work. Category 1 anchors the analysis in transnational circulation and ensures that humour is understood as part of a shared global media ecology rather than a purely Australian invention.

**Category 2: Martian canals and the satire of modern engineering**

This satirical mode is especially evident in repeated comparisons between Martian canals and contemporary engineering projects whose scale, cost and delays were widely reported. Whereas much of this engineering humour continued to circulate internationally, its rhetorical function shifted in Australian newspapers. Unlike the inherited metropolitan wit of Category 1, this mode involved the active deployment of engineering satire by local editors to contextualise and critique contemporary concerns. The humour no longer merely *arrived with* overseas science news; it was increasingly *worked with* and actively redeployed by Australian editors.

Articles increasingly returned to familiar jokes about Martians duplicating the Suez Canal or completing the Panama Canal but now used these tropes less as borrowed wit and more as vehicles for commentary on engineering ambition and failure. Elsewhere, canal systems were quantified in "Suez equivalents" or imagined as eclipsing both imperial showpieces and local colonial works (e.g., *Argus*, 20 August 1888). Such humour drew directly on readers' familiarity with engineering modernity—its ambitions, failures and financial scandals—and redeployed Mars as a detached vantage point from which these terrestrial concerns could be mocked safely. The joke is not about Mars *per se*, but about the hubris and delays of terrestrial engineering culture. The paragraph from the *Argus* cited above exemplifies the notion that Mars was mobilised as a critical mirror for nineteenth-century engineering modernity. It also shows that Australian newspapers were comfortable satirising European projects through planetary analogy.

---

[4] The Mildura irrigation experiment was a highly publicised late-nineteenth-century attempt to transform semi-arid land along the Murray River in Victoria through large-scale canal engineering (for Martian parallels, see also the *Upper Murray and Mitta Herald*, 28 June 1888). The "Chief Secretary" was the senior minister responsible for internal administration and public works. References to his "pet schemes" were a common journalistic shorthand for ambitious but contested state-backed infrastructure and irrigation projects.

Further, the *Australian Star* of 27 April 1895, while reporting on a British Astronomical Association meeting held in Sydney, states, "… the work involved would be equivalent to more than one and a half million Suez Canals". Here, humour operates through absurd numerical scaling. By converting Martian canals into Suez equivalents, the article invokes the canal as a universal unit of engineering labour, implicitly highlighting the implausibility of artificial construction on Mars. This example supports the notion that newspapers used terrestrial engineering projects as interpretive scaffolding. It also bridges Categories 2 and 4 (see Section 6), since numerical excess often tipped from satire into parody.

Such satire recurred across the colonial press. As we saw already, in 1898 Australian newspapers reprinted commentary suggesting that Mars was "… a place where all good Panama promoters might hope to go," explicitly aligning speculative astronomy with the language of finance, promotion and infrastructural optimism. This line—imported but extensively reproduced—uses the Panama Canal not as a marvel but as a symbol of speculative excess, corruption and failure. Here, satire shifts from technological comparison to moral judgement. Elsewhere, jokes about locks, unfinished canals and escalating costs recast Mars as a world in which the perennial problems of terrestrial engineering—budget overruns, technical setbacks and political interference—had been effortlessly overcome, e.g.:

> **(1) Lock-level inference:**
> One canal, communicating with a vast fresh water ocean, has a lock. Possibly it was that circumstance which caused M[onsieur]. de Lesseps to change his mind and apply the wrinkle to his Pacifico–Atlantic ditch. (*Ballarat Star*, 15 August 1888; and many reprints)
>
> **(2) Jokes about unfinished canals:**
> Astronomers have already called the lines going from sea to sea on the surface of Mars 'canals,' but M[onsieur]. [Henri Joseph Anastase] Perrotin [1845–1904; director of Nice Observatory] says that some of these waterways are, like the Panama project, still unfinished. (*Maitland Mercury and Hunter River General Advertiser*, 5 July 1888; and reprints)
>
> **(3) Jokes about escalating costs, finance and budget overruns:**
> Many speculators would like to know what it cost to construct a ship canal in Mars, how many times the capital was augmented before the opening day, if a lottery loan had been resorted to, and what recompense was accorded to the piercer. (*Ballarat Star*, 15 August 1888; and many reprints)

The humour lies in imagining Mars as a refuge for discredited terrestrial engineers. This example strengthens the critical edge of Category 2: satire is directed not at astronomy but at the political economy of engineering modernity. By exaggerating Martian competence, newspapers implicitly criticised Earthly projects whose ambitions outstripped their execution. Similarly, references to duplication on Mars of the Suez Canal—such as the example from the *Argus* of 20 August 1888 cited above—casts Martian canals as already implementing schemes merely *proposed* on Earth. The humour depends on the gap between engineering imagination and execution. This reinforces the idea that Mars allowed newspapers to explore unrealised engineering futures with ironic detachment.

Importantly, this humour was not anti-scientific. It did not deny the existence of canals outright, nor did it reject astronomical authority. Instead, it targeted the broader cultural milieu in which engineering had become a symbol of modern progress. Mars served as a satirical extreme: a planet where engineering rationality appeared to have been perfected to a degree that exposed the fragility of human achievements. In this sense, humour functioned as a form of social commentary embedded within ostensibly scientific reporting.

The international character of this satire is striking. References to Suez, Panama, Russian canal (e.g., *Evening News*, 19 April 1898) schemes and European engineers such as de Lesseps circulated freely through Australian newspapers, demonstrating the extent to which colonial readers were integrated into global conversations about technology and modernity. Martian canals thus became a shared imaginative resource within a transnational media environment. They allowed journalists to combine planetary-scale speculation with the very real uncertainties of nineteenth-century infrastructural capitalism.

Another illustrative use of this category of humour, published in the *Maffra Spectator* of 12 November 1894, a Victorian newspaper, refers to "… a channel indeed which would throw the Suez Canal rather into the shade, and even eclipse the one at Sale!" This example is particularly valuable because it layers scales from the 'global' Suez Canal to the regional Victorian canal at Sale, a recently constructed canal in Victoria's Gippsland region—a modest but much-discussed regional navigation and drainage project. Here, humour is no longer merely international in reference but explicitly local in resonance. By placing Sale alongside Suez, the article collapses planetary speculation into a recognisably Australian frame, signalling a shift from passive reception to local cultural production.

The humour works by collapsing planetary, imperial and local engineering into a single absurd hierarchy. This example shows that engineering satire was not purely metropolitan. Australian newspapers localised global engineering discourse, reinforcing the idea that Mars functioned as a culturally flexible reference point. Local irrigation projects such as the Sale canal and the Mildura schemes form a recurring regional motif in this humour. Both were widely reported, heavily promoted and persistently problematic, making them ideal reference points for comic comparison. By positioning Martian canals as dwarfing these ventures, newspapers transformed familiar colonial struggles with water, labour and scale into a form of planetary satire. Mars thus becomes an imagined space in which the logistical and environmental constraints that plagued Australian irrigation appear effortlessly overcome.

By framing Martian canals through the lens of engineering satire, Australian newspapers transformed speculative astronomy into a vehicle for reflecting on the promises and limits of modern technological ambition. Category 2 thus marks the first stage at which humour became a tool of interpretation rather than transmission. Australian newspapers did not merely reproduce jokes about Martian canals; they used those jokes to interrogate the meanings of progress, competence and modernity in their own world. The humour lay not only in the improbability of canals on Mars but in the uncomfortable suggestion that imaginary Martians might be more capable engineers than their terrestrial counterparts. In this way, jokes about Martian canals reveal less about beliefs in extraterrestrial intelligence than about contemporary anxieties surrounding progress, expertise and the uneven outcomes of global modernity, with humour operating as a form of cultural boundary-work rather than simple disbelief (Gieryn, 1983; Shapin, 1994).

**Category 3: Humour as provisional engagement with speculative astronomy**

Between the light novelty of early Mars reporting and the later excesses of exaggeration and parody lies a distinct intermediate mode: humour used as a means of provisional engagement. Humour emerged not from analogy or exaggeration but from uncertainty itself. Newspapers increasingly joked about the difficulty of seeing clearly, the instability of interpretation and the coexistence of incompatible explanations. Australian newspapers neither treated Mars speculation as mere entertainment nor subjected it to outright ridicule. Here, humour functioned as a response to epistemic fragility rather than to engineering ambition, as a rhetorical device that allowed speculative ideas to be entertained while remaining explicitly unsettled. Humour did not merely trivialise the debate. Instead, it provided a socially acceptable means of acknowledging the speculative nature of the canal hypothesis while continuing to treat astronomy as a legitimate field of inquiry.

Articles in this category typically acknowledged contemporary scientific observations—new surface markings, improved telescopes or favourable oppositions—while framing their implications playfully or conditionally. When the presence or absence of a Martian atmosphere suitable for life support was discussed extensively, Australian papers took the opportunity to involve their local experts, e.g.:

> Mr. [Robert L.J.] Ellery [1827–1908], in fact, inclines to the opinion not that there is no atmosphere in Mars, but that the atmosphere of that planet is of even greater density than that which surrounds the earth." (*Argus*, 21 August 1894; and numerous reprints)

> Mr. Ellery, the Victorian Government astronomer, doubts the genuineness of the recent discovery alleged to have been made as to the absence of an atmosphere in the planet Mars. (*Brisbane Courier*, 27 August 1894)

Rather than drawing humour from engineering analogy or exaggerated scale, Category 3 humour targets the instability of interpretation. Several articles mock the idea that telescopes can reveal implausibly fine detail, implicitly questioning observational claims without rejecting the science of astronomy itself. For instance, in the article in the *Maffra Spectator* we already encountered (12 November 1894), we read, "The most the highest-powered telescope will do is to reduce the distance separating us from Mars to 35,000 miles [56,000 km] …" This passage becomes humorous not through engineering analogy (Category 2), but through optical logic stretched to absurdity. The joke rests on the mismatch between what telescopes can plausibly resolve and what observers claim to see. A second example takes this editorial scepticism to its extreme:

> "Professor [William Henry] Pickering [1858–1938], of Harvard, the American astronomer, asserts that he has discovered 40 small lakes in the planet Mars." That's nothing! Last time our representative was on a visit to this planet he dropped across six seas, fourteen mountains, and a slygrog shop [an unlicenced vendor of alcoholic drinks] besides the lakes aforesaid; but the excursion was not fraught with much pleasure as it followed immediately on top of a banquet, and was mixed with snakes, blue-devils [*delirium tremens*, hallucinations] and onions! (*Nepean Times*, 15 October 1892)

The joke does not dispute Pickering's observational claim directly. Instead, it collapses the distinction between scientific observation and subjective experience, placing telescopic discovery alongside drunken hallucination. The telescope itself, or rather telescopic inference, becomes the object of humour. In fact, the author of the *Argus* article of 20 August 1888 turns this logic on its head, exposing its symmetry through deliberate absurdity:

> Of course there is the further alternative supposition that, even if human life does exist upon the planet Mars, it has had nothing to do with the canals and the great flood that has recently been observed there. And if this opinion prevail[s], we may justly hope that a similar view will be taken by Martian observers of some terrestrial phenomena, and that the ruts on the St. Kilda road,[5] for instance, which must, we imagine, be plainly visible through an ordinary telescope at a distance of only about 40 millions of miles [64 million km], will be attributed to purely natural agencies.

Here, humour arises not from disbelief in astronomy but from the recognition that observational inference, once untethered from clear limits, becomes infinitely reversible. By imagining Martians misinterpreting mundane terrestrial features, the article invites readers to reflect on the fragility and reversibility of telescopic knowledge itself.

The humour targets epistemic limits—how much can really be known through instruments? It invites scepticism without naming any astronomer as fraudulent. In Category 3, humour softened the transition from observation to interpretation, enabling readers to contemplate the possibility of Martian canals or inhabitants without requiring belief in them. The speculative leap was marked as imaginative rather than evidentiary. Several late-1890s articles explicitly joke about the fallibility of observers, sometimes implicitly at the expense of figures such as Percival Lowell (1855–1916), the most prominent American proponent of canal theory: "The structure and the possible inhabitants of Mars form a theme of never[-]ending fruitfulness for astronomical speculators …" (e.g., *Maitland Daily Mercury*, 18 October 1898; and numerous reprints). This framing already signals uncertainty, but humour emerges when elaborate canal systems are presented as speculative constructs layered atop ambiguous visual cues. The humour lies in speculative excess built on fragile perception, using conditional grammar as a device, marking speculation as imaginable rather than demonstrable and keeping interpretation deliberately unresolved. The joke is not that canals are impossible, but that so much confidence rests on such tenuous observation.

---

[5] St. Kilda Road, a major thoroughfare in Melbourne, was already notorious in the late nineteenth century for its rutted surface and poor maintenance.

A particularly strong strand of Category 3 humour comes from juxtaposition, where newspapers place incompatible explanations side by side, e.g., "… opinion is occupied with the presumed canals in the planet Mars" (*Ballarat Star*, 15 August 1888; and numerous reprints). The phrasing "presumed canals" is already distancing, and the subsequent playful speculation we often find in these articles exaggerates interpretive disagreement into humour. A characteristic example follows, which layers technical speculation, journalistic irony and popular imagery within a single passage:

> One canal, communicating with a vast fresh water ocean, has a lock. ... According to the latest astronomical interviewer, Mars would seem to resemble not a little Holland; not only by its canal system, but by its windmills, barge traffic, and winged inhabitants. (*Ballarat Star*, 15 August 1888)

The humour arises from interpretive instability. Multiple explanations coexist, and the article makes that coexistence itself amusing. Some articles explicitly reference disputes among astronomers, then treat those disputes lightly, e.g., "… if these are really the work of living beings —and the regularity of their direction seems an argument against their merely geological origin—their makers must possess high intelligence and engineering skill …" (*Australian Town and Country Journal*, 24 October 1896). The conditional framing is key. The humour is subtle, but it lies in suspending judgement, allowing the reader to enjoy speculation while recognising that no conclusion is secure. This is humour born of uncertainty, not exaggeration. It occupies the space between belief and disbelief. In fact, several articles adopt a tone of conditional irony: "*If* Mars has canals, …", "*If* they are artificial, …", "*If* observers are correct …"

This rhetorical structure appears repeatedly across the 1888–1896 timespan, e.g., "To many people the Panama Canal seems an enterprise as dim and as much out of human reach as the canals which astronomers report in the planet Mars…" (*Shepparton Advertiser*, 19 February 1889). Here, doubt about terrestrial engineering is explicitly mirrored by doubt about astronomical claims, with neither offered as stable ground. The humour hinges on analogy grounded in uncertainty, not on parody or scale. This strategy also helped Australian newspapers manage their relationship to overseas authority. By reproducing metropolitan debates in a lightly humorous or ironic register, colonial editors could relay international scientific discussion while suspending judgment. Humour thus operated as a form of epistemic caution, signalling interest without commitment.

In essence, this mode reflects the unsettled epistemic status of Mars in the late nineteenth century. The planet was framed as a legitimate object of scientific inquiry, but one whose meaning remained unresolved. In this respect, Australian newspapers did not stand apart from metropolitan cultures of scientific scepticism but offer a particularly rich and well-documented example of how humour functioned internationally as a tool for negotiating the limits of astronomical credibility. Humour here did not negate science; rather, it acknowledged uncertainty. Only later, as interpretations became more elaborate, polarised and rhetorically amplified, would humour shift decisively towards exaggeration and sceptical parody.

**From scientific satire to colonial self-positioning**

Taken together, the humorous modes discussed so far reveal a clear transformation in the rhetorical work performed by Martian canals in Australian newspapers. In earlier examples, humour operated primarily as a form of scientific satire. Jokes about optical illusion, over-interpretation and the fallibility of expert observers allowed newspapers to participate in international debates while maintaining a posture of critical distance. The canal hypothesis could be questioned or gently ridiculed without requiring readers to take a definitive position on its validity.

As coverage developed, however, humour increasingly acquired a second function: it became a means of situating Australia within a global landscape of science, technology and modernity. References to European and American astronomers continued to structure authority, but the joke itself was often anchored locally, as when alleged Martian works were said to make the "Mildura experiment" or even "the one at Sale" appear insignificant. Martian canals were no longer merely a problem of astronomical interpretation; they became a mirror through which colonial readers could reflect on their own infrastructural ambitions, environmental challenges and technological limitations.

This transition is significant because it demonstrates how international scientific ideas were not simply imported into the colonial press but reworked in relation to local experience. Humour provided an especially flexible mechanism for this process. It allowed newspapers to align themselves with metropolitan scepticism while simultaneously asserting a distinctively Australian perspective grounded in practical knowledge of land, water and scale. The following section explores this second dimension in greater detail, examining how jokes about Martian canals intersected with colonial self-awareness and infrastructural comparison. In doing so, it shows how Australian newspapers used humour—not only to mediate scientific uncertainty but also to articulate a reflective position within an increasingly globalised scientific culture. In practice, Categories 4 and 5 represent increasing distance from evidentiary claims.

**Category 4: Humour, scale and colonial self-awareness**

In a further stage of humorous engagement, Australian newspapers increasingly treated Martian canals through exaggerated scale, numerical excess and parodic escalation. In this category, humour no longer arose primarily from analogy (Category 2) or uncertainty (Category 3), but from taking speculative claims to their logical extreme. These exaggerated treatments were not distributed evenly across the press, however. They revealed important differences between metropolitan and regional newspapers in how scale and comparison were mobilised, reflecting differences in audience, editorial style and lived experience. While capital-city papers often framed humour through literary allusion or international analogy, regional newspapers tended to ground their jokes in practical comparisons drawn from local infrastructure, irrigation schemes and environmental constraints, because readers had direct experience of such constraints and infrastructural failure.

Metropolitan dailies such as the *Argus*, *The Age*—both Melbourne-based newspapers—and the *Sydney Morning Herald* most frequently employed irony and parody to situate Martian canals within global narratives of technological ambition. Their humour drew explicitly on well-known international projects—most notably the Suez and Panama Canals—and on the reputations of European and American astronomers. Jokes functioned as a form of cultural literacy, signalling that colonial readers were conversant with metropolitan debates and capable of appreciating their excesses. The humour was outward-looking, aligning Australian audiences with a transnational community of scientifically informed readers and positioning colonial newspapers as competent interpreters of metropolitan excess rather than passive recipients of it.

By contrast, regional and provincial papers more often anchored humour in direct, localised comparison. Here, we can draw on an example cited earlier, e.g., "… they are supposed to have carried out an irrigation work by the side of which the Mildura experiment and all the Chief Secretary's pet schemes in addition look very small indeed" (*Argus*, 20 August

1888). Here, the humour exceeds engineering satire. The escalation lies in totalisation, in the sense that Martian engineering absorbs *all* terrestrial schemes and local colonial projects are rendered laughably inadequate by planetary comparison. The humour arises from excessive scope, not analogy. Mars is no longer a mirror; it is an absurd superlative benchmark.

References to Australian irrigation experiments, unfinished public works or notorious engineering challenges translated Martian speculation into familiar terms. Jokes about canals "throwing the Suez into the shade" or eclipsing modest local projects were effective precisely because readers understood the difficulty of constructing large-scale water infrastructure under Australian conditions, giving exaggerated Martian claims a distinctly ironic resonance grounded in lived experience. The *Australian Star*'s statement that "… the work involved would be equivalent to more than one and a half million Suez Canals" (27 April 1895) is a textbook Category 4 example. Unlike Category 2, where Suez operates as an analogy, here it becomes a quantitative unit pushed beyond intelligibility, directly reflecting experiential knowledge. The number is so large that comparison collapses into numerical overload: scientific precision itself becomes the source of humour. The humour lies in the collapse of scale: calculation becomes parody. This example is particularly valuable because it shows humour emerging from within apparently serious scientific reasoning, rather than being imposed externally.

Examples such as the following quote from the *Geelong Advertiser* of 13 March 1897 mark a shift from sceptical humour to parodic awe: "… compared with which our Suez and Manchester Canals are absolutely insignificant." Here, the Manchester Ship Canal was invoked because it was a culturally saturated unit of engineering achievement—recent, costly, rational and widely understood. Comparing Martian canals to it allowed journalists to move seamlessly from scientific speculation into parody, using a familiar terrestrial benchmark to expose the implausibility of planetary-scale engineering claims. The language ("absolutely insignificant") is deliberately extreme. This is a clear example of Category 4 humour in that there is no critical distance or irony aimed at Earthly projects. Instead, Martian engineering is inflated to a mythic level. The exaggeration destabilises the claim rather than endorsing it: Mars becomes a caricature of perfect rational planning—too perfect to be credible.

Taken to extremes, another such example of (Lowell-inspired) exaggeration is found in the *Riverine Herald* of 26 October 1896: "… a complicated scheme of irrigation extended over the entire area of a planet … compared with it, such pieces of work as the Forth Bridge[6] or the Suez Canal are almost infinitely insignificant." Here the exaggeration is spatial and systemic; not a canal, not a network, but the entire planet. The phrase "almost infinitely insignificant" is doing overt comic work. This example is important because it shows how Lowellian arguments were received parodically even when reported respectfully. The article takes Lowell's idea of rational, large-scale Martian engineering and blows it out of proportion, imagining irrigation over the *entire planet*, not just canals. The humour comes from the mismatch between Lowell's scientific claims (reported seriously) and the absurdity of its implications when imagined literally.

We also see familiar examples of earlier humour re-emerge. For instance, where Mars is referred to as "… a place where all good Panama promoters might hope to go" (*Evening News*, 17 October 1898; citing the *Pall Mall Gazette*) in Category 1, this is imported humour. In Category 4, its repetition and redeployment matter. By 1898, this line circulates so widely that it becomes almost a stock joke—shorthand for speculative excess. By this time, the humour is no longer contingent on Panama alone. Mars itself has become a parodic space for implausible ambition. The canal hypothesis functions as a narrative engine for comic escalation.

Finally, Category 4 humour is found in phrases such as "… windmills, barge traffic, and winged inhabitants" (*Ballarat Star*, 15 August 1888; and numerous reprints). This is a clear case of "parodic accretion": each added detail increases implausibility. The article does not stop at canals but keeps layering imagery. This distinguishes Category 4 from Category 3: Category 3 jokes about uncertainty, whereas Category 4 ridicules overdetermination—too many explanations, too much structure. This shows that humour operated as a mediating layer between global scientific discourse and local material reality. Even when drawing on identical international reference points, metropolitan and regional papers mobilised them differently. Humour allowed Australian newspapers to engage actively with overseas scientific speculation while asserting interpretive agency. The canal debate thus became a site where international astronomy was refracted through distinctly colonial understandings of scale, labour and environment.

---

[6] The Forth Bridge (Scotland) was opened 1890. At the time, it was a marvel of engineering and an iconic symbol of human achievement. Its construction involved enormous scale and technical difficulty.

**Category 5: Humour as scepticism and boundary-setting**

Alongside parody and scale-based exaggeration, humour in Australian Mars reporting also functioned as a mechanism of scepticism and boundary-setting, marking the limits between plausible scientific interpretation and speculative excess. Rather than merely entertaining readers, jokes about Martian canals frequently served to mark the boundaries of credible scientific interpretation. By ridiculing the more extravagant implications of canal theory, newspapers could acknowledge popular fascination with Mars while simultaneously signalling distance from its most speculative claims. This form of humour was often directed less at the putative existence of linear markings than at the interpretive leap from observation to civilisation. Jokes about Martian engineers, planetary irrigation boards or interplanetary commerce allowed newspapers to rehearse Lowellian ideas in exaggerated form, thereby exposing their implausibility without the need for explicit technical rebuttal. Satire thus offered an economical way to express doubt in a media environment where direct criticism of prominent scientific figures might otherwise appear presumptuous.

 Category 5 humour marks the point at which Martian canals cease to function as a scientific hypothesis or even a satirical analogue. Instead, Mars becomes a fully fictionalised cultural stage, populated by engineers, republicans, winged beings and failed financiers. Humour no longer mediates uncertainty (Category 3) or exaggerates scale (Category 4) but detaches entirely from evidentiary concerns. The canal system survives only as narrative scaffolding for social, political and economic parody. Importantly, this sceptical humour aligned Australian newspapers with a broader international undercurrent of resistance to canal literalism. British and American periodicals frequently employed similar strategies, and Australian papers often reprinted or adapted these pieces, embedding local readers within a transnational culture of amused caution. Humour therefore acted as a shared rhetorical resource through which colonial audiences could participate in global scientific debate while avoiding uncritical endorsement. Mars is no longer a mirror, analogy or exaggerated benchmark, but it has transitioned into a fully comic world, used for free-ranging parody, political satire and cultural play. The canal hypothesis becomes a pretext rather than the point.

 The well-rehearsed line attributed to the *Pall Mall Gazette*, where "… the planet [Mars] is one vast network of canals, a place, in fact, where all good Panama promoters might hope to go" had effectively transitioned from Category 1 (imported wit) to Category 5 by the late 1890s. Its repeated circulation strips it of topical specificity and turns it into stock parody. The joke no longer depends on belief or disbelief in canals; it assumes canals as narrative infrastructure for satire. Mars is now a fully fictionalised social space, no longer a scientific problem. Even Lowell-inspired descriptions, when reprinted, could function parodically rather than persuasively. The tone shifts from sceptical scaling to mythic awe bordering on self-parody: "He describes them as 'uncanny' in their aspect … compared with which our Suez and Manchester Canals are absolutely insignificant" (*Geelong Advertiser*, 13 March 1897). The word "uncanny" signals that the article is no longer evaluating claims but enjoying their strangeness. Mars is elevated into a quasi-legendary realm of superhuman competence.

 Similarly, the point that "Mars would seem to resemble not a little Holland; not only by its canal system, but by its windmills, barge traffic, and winged inhabitants" (*Ballarat Star*, 15 August 1888) is no longer exaggeration of scale (Category 4), but parodic accretion. Each added detail moves further from astronomy and deeper into comic invention. "Winged inhabitants" reflects a deliberate abandonment of evidentiary restraint. Mars becomes a theatrical setting, enabling whimsical elaboration unconstrained by scientific plausibility. In other publications, Mars is explicitly politicised: "One speculator concludes from the uniformity in the height of the mountains, the width of the rivers, and the size of the ships, that Mars must be a republic" (*Ballarat Star*, 15 August 1888; and reprints):

> Who knows but that the three classical incarnations of Father Mars might not be after all those tutelary deities — Liberty, Equality, and Fraternity. And since the wolf and the woodpecker were sacred to the bellicose god, do not these symbolise some republics whose devotees are constantly devouring and hammering one another?

Astronomical features are pressed into service as evidence for a comic theory of Martian governance. In the *Illustrated Sydney News* of 18 February 1893, we encounter a similarly absurdist passage, following a lengthy scientific explanation:

> … if the Martial [*sic*] Government is not subject to socialistic interference or monetary deficits, it may have voted the necessary money to establish a trans-celestial signalling station or balloon away, and the Martians will thereby be enabled to enlighten us on all the doubtful points themselves.

This is not uncertainty (Category 3) or exaggeration (Category 4), but ideological parody, using Mars to satirise political rationalism and republican ideals on Earth. Taking this to its extreme, some commentators turned Mars into a bureaucratic comedy:

> Many speculators would like to know what it cost to construct a ship canal in Mars, how many times the capital was augmented before the opening day, if a lottery loan had been resorted to, and what recompense was accorded to the piercer. (*Ballarat Star*, 15 August 1888; and reprints)

This passage fully detaches Mars from astronomy. The language of prospectuses, cost overruns and public finance is applied absurdly to an imaginary planet. The canal hypothesis is now a narrative prop, not an object of inquiry. Finally, Mars is now frequently used as an imagined escape from Earthly limits, e.g., "To many people the Panama Canal seems an enterprise as dim and as much out of human reach as the canals which astronomers report in the planet Mars …" (*Shepparton Advertiser*, 19 February 1889). The joke works by collapsing two unrealities into one another. Mars becomes shorthand for impossibility itself. The humour no longer interrogates astronomy; it uses Mars as a metaphor for human overreach. An excellent example of this framing is found in *The Age* of 1 December 1894:

> Strange coincidence, the new Panama Company has just recommended the "completion" of the interocean ditch. By the time the latter is achieved M[onsieur] Nicolas Camille]. [de] Flammarion [1842–1925; French astronomer] … shall have established electrical communication between ourselves and the sovereign people in Mars.

Here, Mars functions not as a speculative scientific object or topical satire but as a rhetorical shorthand for the unattainable future, a comic register of impossibility against which terrestrial ambitions are measured. Combined with the *Shepparton Advertiser* example cited above, the Panama-promoters joke and the bureaucratic-cost parody, this example confirms that by the mid–late 1890s Mars had become a stable metaphor for fantasy itself, no longer requiring any astronomical scaffolding.

At the same time, sceptical humour reinforced the authority of observation over interpretation. By laughing at speculative excess, newspapers implicitly reaffirmed the values of restraint, measurement and methodological care that underpinned earlier Mars reporting. The joke became a boundary marker: curiosity was acceptable, imagination tolerated, but claims of intelligent design were pushed into the realm of entertainment rather than science. Seen in this light, humour did not undermine scientific seriousness; it helped preserve it.

Through irony and satire, Australian newspapers negotiated their position between fascination and disbelief, aligning themselves with international norms of scientific caution while maintaining an accessible and engaging mode of presentation.

**Synthesis: Humour as a dynamic repertoire**

Taken together, Categories 2 through 5 demonstrate that humour in Australian Mars reporting was not a single response to scientific uncertainty but a dynamic ecosystem that evolved alongside the international debate.

In Category 2, humour functioned primarily as a tool of accessibility. Light jokes, whimsical comparisons and playful language rendered astronomical material engaging for general readers without challenging its authority. Mars was interesting, distant and novel; humour served to domesticate science rather than interrogate it.

Category 3 marks a shift from accessibility to provisionality. As observational claims accumulated and interpretive stakes increased, humour became a way to manage uncertainty. Speculative ideas were entertained but hedged. This form of humour acknowledged scientific debate without resolving it, allowing Australian newspapers to remain participants rather than arbiters of international controversy.

Category 4 represents a further escalation, in which humour exaggerated Martian claims to the point of absurdity. Here, the scale, regularity and engineering implications of the canals were amplified through comparison with terrestrial megaprojects. Exaggeration did not necessarily reject canal theory, but it rendered its implications visible—and questionable—by stretching them beyond plausibility.

Finally, Category 5 marks the consolidation of humour as sceptical commentary. In this mode, exaggeration gave way to parody, irony and outright mockery. Humour functioned explicitly as a boundary-setting device, signalling that speculative interpretations had exceeded acceptable scientific limits. Mars became not just an object of wonder, but a site where scientific credibility itself was contested.

Seen in sequence, these categories trace a rhetorical progression rather than a typology of static responses. Australian newspapers adapted their humour as Mars moved from novelty to problem to controversy. This progression mirrors the broader international trajectory of Mars discourse: starting with early astronomical reports of linear features on Mars during the polanet's 1877 opposition by Giovanni Vrginio Schiaparelli (1835–1910), whose *canali* were widely (if mistranslated) taken as evidence of surface channels, sparking global interest in the possibility of extraterrestrial life; continuing through the late-nineteenth and early-twentieth-century promotion and serious debate over Martian 'canals' by figures such as Percival Lowell and commentators across Europe and North America; and culminating in heated public and scientific controversy as scepticism about the reality of canals grew and alternative interpretations were advanced (see also de Grijs, 2026a). This demonstrates that colonial journalism was not merely reactive but rhetorically agile within a global scientific conversation.

**Conclusion**

Humour in late nineteenth-century Australian Mars reporting was neither accidental nor uniform. It formed part of a dynamic rhetorical ecosystem that allowed newspapers to engage with scientific uncertainty, speculative excess and international authority without committing to fixed interpretive positions. By tracing five successive humorous categories, I have attempted to show how Australian journalism adapted its tone as Mars shifted from a distant astronomical object to a contested site of scientific meaning.

Although humorous commentary on the canals of Mars appeared in newspapers elsewhere, the Australian press displayed a particularly sustained and distinctive use of comic framing. In a colonial society accustomed to satire and sceptical commentary in public life, humour provided a culturally familiar way of engaging with extraordinary astronomical claims.

In its earliest form, humour entered Australian reporting through the reproduction of metropolitan material, embedding colonial readers within a transnational media ecology. As local coverage expanded, humour became a tool of accessibility, rendering complex astronomical ideas intelligible and engaging. With the rise of canal theory and debates over habitability, humour assumed a provisional role, allowing speculation to be entertained while remaining unsettled. Later, exaggeration and parody exposed the logical and material implications of canal narratives and, finally, humour functioned as sceptical boundary-work, signalling the limits of acceptable scientific inference. This progression mirrors the broader international trajectory of Mars discourse while revealing the active role of colonial newspapers in mediating it. Australian journalism did not simply transmit scientific claims; it calibrated their meaning through tone, irony and scale. Humour thus served as a means of participation rather than detachment, enabling engagement without epistemic overcommitment.

By focusing on humour as a serious analytical category, this paper shows how comic framing allowed newspaper audiences to engage with speculative astronomical claims while maintaining the authority of scientific observation. It suggests that humour deserves closer attention as a mediator of scientific authority, particularly in periods of uncertainty and debate. More generally, it demonstrates that Australian newspapers played a rhetorically sophisticated role in shaping how global scientific controversies were understood at the local level. Future work might extend this analysis beyond Mars to other contested astronomical objects, testing whether similar humorous oeuvres emerged elsewhere.

The Australian press coverage of Mars in the late nineteenth century therefore illustrates more than a local curiosity about planetary science. It reveals how speculative scientific claims could be incorporated into everyday public discourse through culturally familiar rhetorical strategies. Humour allowed newspapers to acknowledge the extraordinary implications of the canal hypothesis while maintaining a degree of critical distance from it. Rather than rejecting astronomy outright or uncritically embracing sensational claims about extraterrestrial life, journalists and readers navigated uncertainty through irony, exaggeration and playful commentary.

The Australian case highlights a broader dynamic in the public life of science: humour can function as a cultural mechanism for negotiating the tension between scientific authority and speculative interpretation. Similar tensions between scientific authority and imaginative speculation continue to characterise public discussions of astronomy and the possibility of life beyond Earth today. In this sense, humour in Mars reporting functioned not simply as entertainment but as a practical means by which readers could acknowledge the speculative character of planetary interpretation while continuing to recognise the authority of astronomical observation. The treatment of Mars in colonial newspapers thus offers a window into how popular audiences engaged with emerging scientific debates during a period when the boundaries between observation, interpretation and imagination remained highly visible.

## References

*Argus* (1888, 20 August). P6.
*Argus* (1894, 21 August). Mr. Ellery on the atmosphere of Mars, 4.
*Australian Star* (1895, 27 April). British Astronomical Association. Notes upon Mars, 2.
*Australian Town and Country Journal* (1896, 24 October). Within narrow bounds, 19.
*Australische Zeitung* (1882, 3 October 1882). Vermischtes, 12. (In German.)
*Ballarat Star* (1888, 15 August). Parisian echoes, 4.
*Brisbane Courier* (1894, 27 August). P4.
Crossley, R. (2011). *Imagining Mars. A literary history*. Wesleyan University Press.
Crowe, M. J. (1986). *The extraterrestrial life debate, 1750–1900: The idea of a plurality of worlds from Kant to Lowell*. Cambridge University Press.
de Grijs R. (2026a). Mars in the Australian press, 1875–1899. 1. Interpretation, authority and planetary science. *Journal of Astronomical History and Heritage*, submitted.

de Grijs R. (2026b). Mars in the Australian press, 1875–1899. 2. Circulation and attribution. *Journal of Astronomical History and Heritage*, submitted.
*Evening News* (1898, 17 October). P.7.
*Evening News* (1898, 19 April). P4.
*Geelong Advertiser* (1897, 13 March). The planet Mars. Is it inhabited? P4.
Gieryn, T. F. (1983). Boundary-work and the demarcation of science from non-science: Strains and interests in professional ideologies of scientists. *American Sociological Review*, *48*(6), 781–795.
*Illustrated Sydney News* (1893, 18 February). An interesting celestial neighbor, Mars, 18.
Lane, K. M. D. (2011). *Geographies of Mars. Seeing and knowing the red planet*. The University of Chicago Press.
Lightman, B. (2007). *Victorian popularizers of science. Designing nature for new audiences*. The University of Chicago Press.
*Maffra Spectator* (1894, 12 November). Mars, 3.
*Maitland Daily Mercury* (1898, 18 October). Mars, 2.
*Maitland Mercury and Hunter River General Advertiser* (1888, 5 July). Suez mail extracts, 6.
Moyal, A. (1984). *Clear across Australia: A history of telecommunications*. Thomas Nelson. https://archive.org/details/clearacrossaustr0000moya.
Potter, S. J. (2012). *News and the British world: The emergence of an imperial press system, 1876–1922*. Clarendon.
*Riverine Herald* (1896, 26 October). Is there another world? P3.
Secord, J. A. (2004). Knowledge in transit. *Isis*, *95*(4), 654–672.
Shapin, S. (1994). *A social history of truth. Civility and science in seventeenth-century England*. The University of Chicago Press.
Sheehan, W. (1996). *The planet Mars: A history of observation and discovery*. University of Arizona Press.
*Shepparton Advertiser* (1889, 19 February). The Panama Canal, 2.
*The Age* (1894, 1 December). Our Paris letter, 9.
*Upper Murray and Mitta Herald* (1888, 28 June). Melbourne gossip, 3.